# Estimating primary energy of cosmic rays by calculating secondary particles density in optimum distance from shower core

G. Rastegarzadeh[1,*] and L.Rafezi[1]

[1] Physics Faculty, Semnan University, Semnan, Iran
*corresponding author; grastegar@semnan.ac.ir

**Abstract**

Optimum distance ($R_{opt}$) is a distance from the shower core in which the density calculated by lateral distribution function, has its minimum uncertainty. In this paper, Using CORSIKA code, proton, carbon and iron primary in the energy range between $10^{13}$ - $3\times10^{15}$ eV are simulated to find $R_{opt}$ for Alborz-I array located at an altitude of 1200 m above sea level.

It is shown that $R_{opt}$ is approximately independent of characteristics of primary particle and it is only dependent to array configuration. Dependency of $R_{opt}$ on layout and detector spacing for 20 Alborz-I array detectors, are studied. It is shown that the Alborz-I array layout and its detector spacing result into the best (minimum uncertainty) $R_{opt}$ for its number of detectors. In this work $R_{opt}$ for Alborz-I array is obtained about 9±1 m. In addition, it is shown that, by finding dependency of primary energy to density in optimum distance, energy of primary particle can be estimated well. An energy estimation function is suggested and the function is examined by another set of simulated showers.

**Introduction**

When high energy cosmic ray ($E > 10^{13}$ eV) interacted with atmosphere molecules, a shower of secondary particles is created called Extensive air shower (EAS). Mass, energy and arrival direction of cosmic rays can be determined by using different characteristics of EAS particles like maximum depth in atmosphere $X_m$ and its fluctuations $\sigma(X_m)$[1], steepness of lateral distribution of particles [2], multiplicity, density and type of secondary particles [3], shower size[4], arrival time [5] etc.

Using ground based array of particle detectors is one of the most common methods to study EAS characteristics by sampling EAS particles. In most cases, the covering surface of array is 1% of total shower surface [6]. Due to the economic and environmental issues, there are limitations in the area of arrays and detectors dimensions. To overcome these limitations a proper lateral distribution function (LDF) should be defined in order to estimate shower size. Uncertainty in the form of the LDF, the position of the shower core and integrating LDF on total area can lead to a noticeable uncertainty in calculating shower size.

Hillas [7] introduced an optimum distance from the shower core ($R_{opt}$), where the uncertainty caused by selecting different LDF models has its minimum value and this point is the best position for measuring density. $R_{opt}$ is a function of the array configuration and each observatory needs to have it before data analyzing. For example, $R_{opt}$ is about 1000 m for a vast observatory like Peirre Auger with 1500 m distance between detectors [8].

Alborz-I array is located at Sharif university of technology, Tehran ( 35° 43' N 51° 21' E), 1200 m a.s.l and it is supposed to detect cosmic rays around the knee region. This array is in construction phase and consist of 20 scintillation detectors (each with surface area of 50×50 cm$^2$) is placed in 1600 m$^2$ area. Array configuration and dimensions is shown in Fig.1 [9].

To determine $R_{opt}$ for Alborz-I array and study the effect of selecting different LDF models on $R_{opt}$, CORSIKA code (version 74xxx) is used to simulate EASs. QGSJET-II and GHEISHA models are used for high and low energy hadronic interactions respectively. In this paper it is shown that $R_{opt}$ is independent of direction, energy and mass of primary particle but not from array configuration. So, as $R_{opt}$ is depend on array configuration, the result of different layouts and detector spacing is presented. Finally, using particle density at $R_{opt}$, an estimation function for calculating primary energy is proposed.

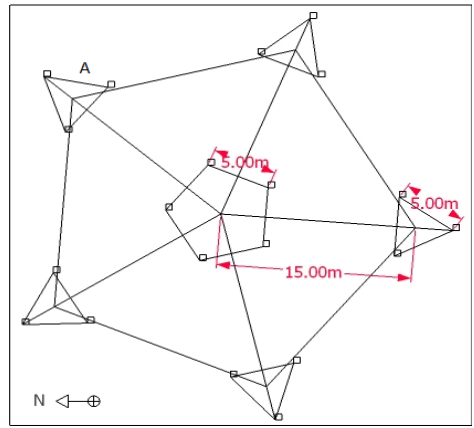

Fig.1: Configuration of Alborz-I array

It should be noticed that for each step of this study 1000 showers were simulated and for each of them 81 different core position (between 0 to 30m) from center of array is chosen for an individual shower which means number of showers is multiplied by 81.

**Calculating $R_{opt}$ for Alborz-I Array**

As it mentioned before, LDF has a crucial role in estimating shower size, finding core location and determining particle density in different distances from the core. In order to specify $R_{opt}$ and obtain density in $R_{opt}$, three conventional lateral distribution functions were used: Power law function

$$\rho(r) = kr^{-\beta} \qquad (1)$$

Haverah Park function [10]

$$\rho(r) = kr^{-\left(\beta + \frac{r}{4000}\right)} \qquad (2)$$

and NKG type function [11]

$$\rho(r) = k\left(\frac{r}{r_s}\right)^{-\beta}\left(1 + \frac{r}{r_s}\right)^{-\beta} \qquad (3)$$

where r is the distance from the shower core, k is a size parameter, β is a slope parameter and $r_s$ is a scale parameter. Similar to β and k, $r_s$ should be a fit parameter but because of its dependence to β, fitting process faces some complications and the common solution is to predetermine $r_s$. Here $r_s$ is assumed 5 times $r_m$ where $r_m$ is Moliere radius [11]. Moliere radius for Tehran altitude (1200 m) is about 95m [12] so $r_s$ would be 475m.

For finding k and β in mentioned LDFs, core position must be given. On the other hand, core position cannot be calculated without having k and β. To overcome this contrast, fit parameters and core position are computed simultaneously by least square method using those functions which introduced above.

Alborz-I array is simulated and based on the previous studies [9] it is preferred to record a shower when all 5 detectors of central cluster are triggered. Fig.2 shows density as a function of core position reconstructed by NKG type function with different slop parameters (β =0.8, β=1.2 & β=1.8) for a vertical 300TeV proton initiated shower. As it is clear in this figure, in a certain point, the effect of β on density measurement becomes minimum, this distance is called $R_{opt}$.

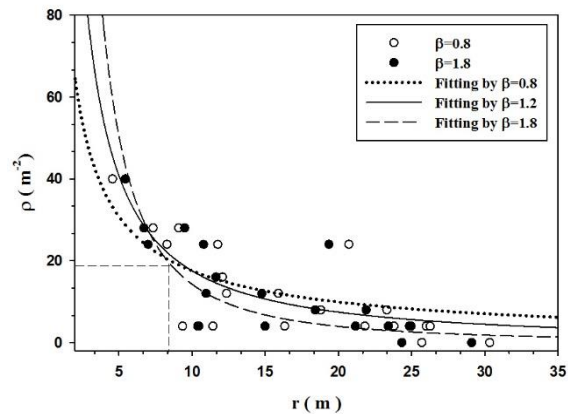

Fig.2: Particle density in detectors versus distance from the reconstructed core position with β =0.8, 1.8. For 3 different slop parameters, fitted to NKG type function, is plotted. Dotted line for β =0.8, dashed line for1.8 and continues line for the intermediate value of β =1.2 (points for β =1.2 are not plotted).

For more precise calculation of $R_{opt}$, numerical and analytical approaches are used. In the former approach, density is computed as a function of r by means of NKG type function repetitively with 50 different β for a single shower. To describe precisely, r changes from 0 to 30m from the shower core with steps of 0.2m and for each step ρ is calculated for 50 different β. The maximum and minimum values of ρ in each step are called $ρ_{max}$ and $ρ_{min}$ respectively. The relative difference of $ρ_{max}$ and $ρ_{min}$ for all steps of r, are obtained and the r with the minimum relative difference is considered as $R_{opt}$. For a vertical proton initiated shower with the primary energy of 300TeV, the relative difference of $ρ_{max}$ and $ρ_{min}$ with respect to r is plotted in Fig.3. As it can be seen in this figure, at the $R_{opt}$ the relative difference value to selected β is minimum. The higher amounts of this difference before and after the $R_{opt}$ are due to the large intrinsic fluctuation near the shower core and low particle density at far distance, respectively.

In analytical method, by minimizing dρ/dβ, an equation for calculating $R_{opt}$ is found [8]. This equation for power law and Haverah Park functions is:

$$\frac{d(\ln k)}{d\beta} = \ln R_{opt} \qquad (4)$$

And for NKG type function is:

$$\frac{R_{opt}}{r_s} = \frac{-1+\sqrt{1+4e^a}}{2} \qquad (5)$$

Where $\alpha = \frac{d(lnk)}{d\beta}$. In order to find $R_{opt}$ for all LDFs, $\frac{d(lnk)}{d\beta}$ needs to be calculated. In order to calculate $\alpha$ for a single shower, $lnk$ is plotted as function of 50 different $\beta$s and the slope of the plot is substituted as $\alpha$ in equations 4 and 5 to obtain $R_{opt}$.

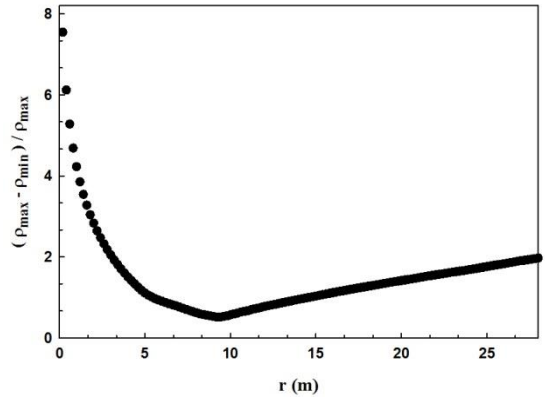

Fig.3: Relative difference of $\rho_{Max}$ and $\rho_{min}$ versus r for one vertical proton initiated shower with the primary energy of 300Tev using 50 different $\beta$.

To compare results of the two methods, $R_{opt}$ calculated using NKG function after imposing mentioned trigger condition for 81000 vertical proton initiated shower with the energy of 300TeV. Distributions of $R_{opt}$ obtained from both methods are shown in Fig.4 and $R_{opt}$ is found about 9 m in both methods. It can be inferred from this figure that results of both methods are completely compatible. Considering this compatibility, only analytical method will be applied in the following.

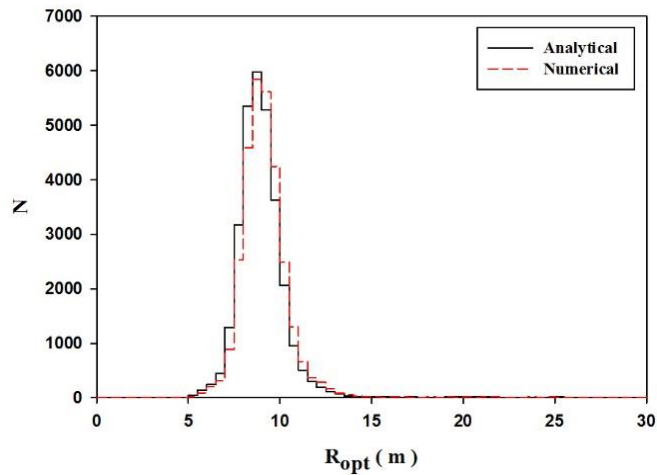

Fig.4: Distribution of $R_{opt}$ from numerical and analytical method.

To investigate the dependence of $R_{opt}$ to LDF model, average $R_{opt}$ is calculated for 81000 vertical proton initiated showers with energy of 300TeV using 3 LDFs and the result is shown in table 1. Results implies that $R_{opt}$, its uncertainty ($\sigma_{R_{opt}}$) and normalized density value in $R_{opt}$ are independent of the LDF model so only NKG type function will be used in the following.

Table 1: Average $R_{opt}$ is calculated for 81000 vertical proton initiated showers with energy of 300TeV using 3 different LDFs

| LDF | $\bar{R}_{opt}(m)$ | $\sigma_{R_{opt}}(m)$ | $\rho_{LDF}(9)/\rho_{NKG}(9)$ |
|---|---|---|---|
| NKG type | 9.07 | 0.96 | 1.0000 |
| Haverah Park | 9.22 | 1.00 | 0.9982 |
| Power law | 9.27 | 0.97 | 0.9965 |

## Dependence of $R_{opt}$ to primary mass, energy and zenith angle

To investigate the dependence of $R_{opt}$ to different primary characteristics, for each Primary particle, Energy and zenith angle, 81000 showers are used. In Fig.5, $R_{opt}$ with versus primary energy and zenith angle is shown. This diagrams show that $R_{opt}$ is independent of primary energy as well as zenith angle.

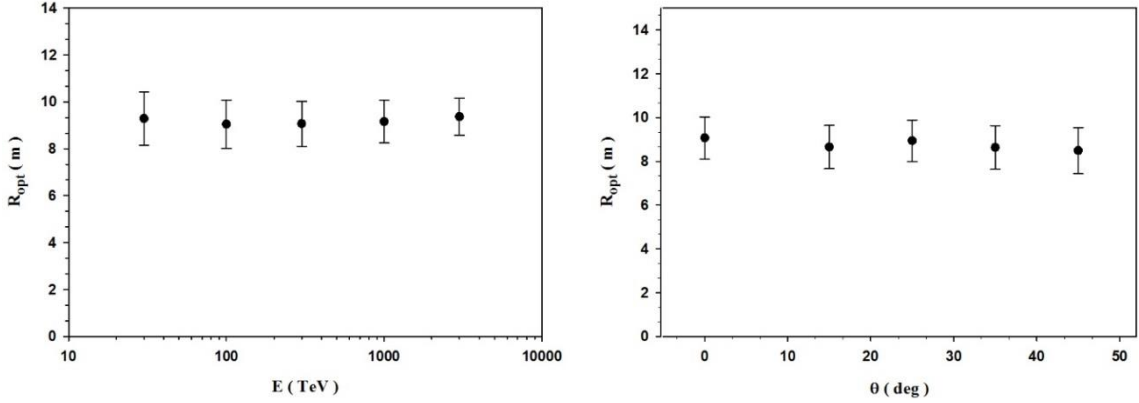

Fig.5: Energy dependence (left) and zenith angle dipendence(right) of $R_{opt}$.

We define trigger probability as:

$$P_{tr} = \frac{N_{trigger}}{N} \quad (6)$$

Where $N_{trigger}$ is the number of showers that fulfill the trigger condition and N is the total number of showers (81000). $R_{opt}$, $\sigma_{R_{opt}}$ and Trigger probability for different primary particles are reported in table 2 for 300 and 3000 TeV primary energies.

Considering 300 TeV primary energy, Iron has lower trigger probability relative to proton and carbon as a result of lower number of secondary particles at observation level. At 3000 TeV energy, but, even Iron primary, has significant $P_{tr}$. Furthermore, Results implies that $R_{opt}$ is independent of primary particle mass.

Table 1: Average $R_{opt}$, $\sigma_{R_{opt}}$ and $P_{tr}$ are calculated using 81000 vertical showers with energy $3\times10^{14}$ and $3\times10^{15}$ eV for 3 different Primary mass.

| E (TeV) | Primary Particle | $\bar{R}_{opt}(m)$ | $\sigma_{R_{opt}}(m)$ | $P_{tr}$ |
|---|---|---|---|---|
| 300 | P | 9.07 | 0.96 | 0.373 |
|  | C | 8.18 | 0.95 | 0.148 |
|  | Fe | 7.92 | 0.97 | 0.026 |
| 3000 | P | 9.37 | 0.78 | 0.998 |
|  | C | 8.80 | 0.78 | 0.994 |
|  | Fe | 8.30 | 0.81 | 0.978 |

## Effect of array configuration on $R_{opt}$

In this section the effects of array layout and distance of the adjacent detectors are discussed using 81000, 300TeV proton initiated showers. At the first step, 4 different layouts (demonstrated in Fig.6) in 1600m$^2$ area are designed and $R_{opt}$ is computed with explained analytical method for all of them. In addition, for this particular study, trigger condition is changed into 3 adjacent detectors for all layouts. It should be noted that in all layouts number of detectors remains 20 and all detectors have 50×50 cm$^2$ dimensions.

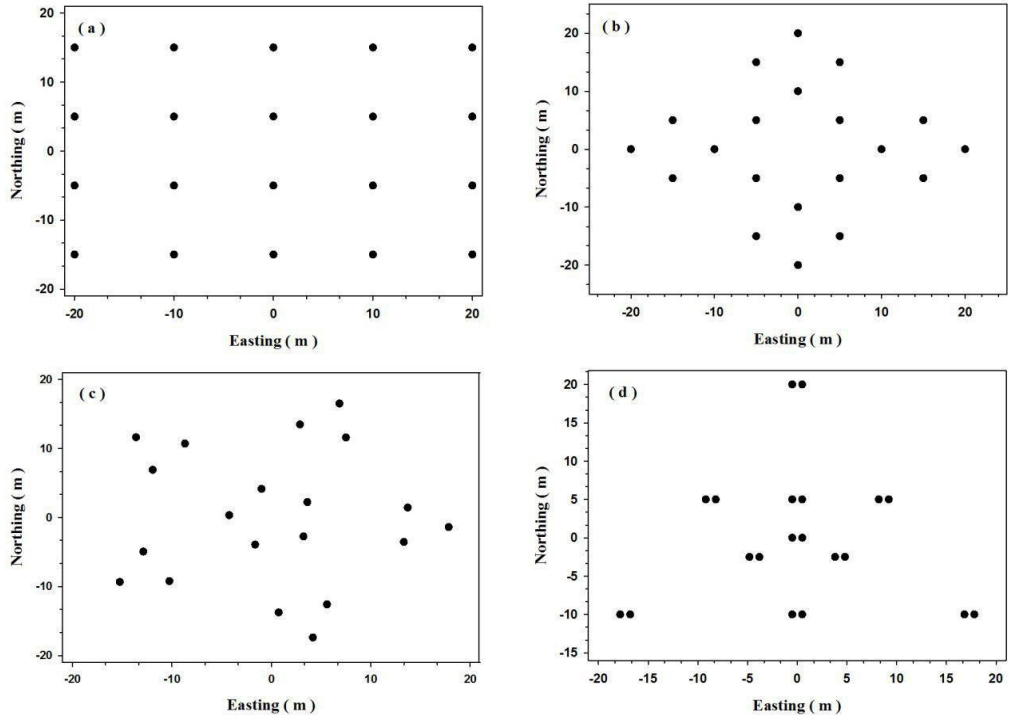

Fig.6: 4 different layouts for 20 scintillation detectors in 1600m$^2$. Layout (c) is desinged layout for Alborz-I array.

Results are compared in table 3 where $d_{min}$ is the minimum distance of two detectors. It is inferred that as $d_{min}$ decrease (detectors get closer to each other) , $R_{opt}$ and $\sigma_{R_{opt}}$ dicresees except for layout (d). It worth to mention that in layout (d), two detectors with 1m distance placed on the vertexes of triangles to check how $R_{opt}$ precision ($\sigma_{R_{opt}}$)is affected by decreasing detector distance extremely. As it can be seen this extreme change has a reverse effect. Table 3, also shows that Alborz-I layout ( layout c) has the best accuracy in finding $R_{opt}$.

Table 3: Average $R_{opt}$, $\sigma_{R_{opt}}$ and $P_{tr}$ are calculated using 81000 vertical proton initiated showers for different layouts with different $d_{min}$

| Array layout | $d_{min}(m)$ | $\overline{R}_{opt}(m)$ | $\sigma_{R_{opt}}(m)$ | $P_{tr}$ |
|---|---|---|---|---|
| a | 10.00 | 14.31 | 2.59 | 0.955 |
| b | 7.07 | 11.02 | 2.05 | 0.933 |
| c | 5.00 | 8.98 | 1.86 | 0.868 |
| d | 1.00 | 9.19 | 2.98 | 0.830 |

In the next step the effect of $d_{min}$ on $R_{opt}$ and its precision is studied for Alborz-I layout, considering trigger condition of 5 central detectors. For this reason array dimensions were changed with scale factors written in table 4. It is forgone conclusion that by changing $d_{min}$, trigger probability ($P_{tr}$) will change.

Scale factor, $d_{min}$, $R_{opt}$, $\sigma_{R_{opt}}$ and trigger probability are shown in table 4. Results show that by increasing the scale factor and hence increasing detector distances, $P_{tr}$ reduces due to decreasing of the density. Moreover with enhancing the scale factor, the number of recorded showers reduces and consequently $\sigma_{R_{opt}}$ increases. Although for smaller scale factors $P_{tr}$ is higher, precision of density estimation decreases. This is because in distances close to the shower core, density estimation is more influenced by intrinsic fluctuation of showers. So it can be said that the Alborz-I layout (scale factor =1) is the best choice. $R_{opt} \cong 9 \pm 1$m for Alborz-I array and trigger condition of 5 central detectors. Comparing tables 3 and 4, it can be concluded that $R_{opt}$ depends on both layout and detector spacing.

Table 4: Average $R_{opt}$, $\sigma_{R_{opt}}$ and $P_{tr}$ are calculated using 81000 vertical proton initiated showers for Alborz-I array layout with different scale factors applied for array dimensions.

| Scale factor | 0.50 | 0.75 | 1.00 | 2.00 | 3.00 | 4.00 | 5.00 |
|---|---|---|---|---|---|---|---|
| $d_{min}(m)$ | 2.50 | 3.75 | 5.00 | 10.00 | 15.00 | 20.00 | 25.00 |
| $\overline{R}_{opt}(m)$ | 4.33 | 6.62 | 9.07 | 19.57 | 29.62 | 41.46 | 56.48 |
| $\sigma_{R_{opt}}(m)$ | 0.49 | 0.74 | 0.96 | 2.42 | 6.36 | 9.83 | 13.49 |
| $P_{tr}$ | 0.40 | 0.39 | 0.37 | 0.30 | 0.22 | 0.15 | 0.09 |

## Energy estimation function

In this section an equation has found to estimate primary energy in terms of calculated density in $R_{opt}$ for Alborz-I array. First, density in $R_{opt}$ ($\rho(R_{opt})$) is calculated for proton and Iron initiated showers which fulfill trigger condition. This is done for 5 different primary energies. Showers are vertical, number of showers for each element is 81000 and energy range is $3 \times 10^{13}$ to $3 \times 10^{15}$eV. Next, LogE versus Log$\rho(R_{opt})$ for both Primary mass are plotted in Fig.7 and an exponential function is fitted to plots .

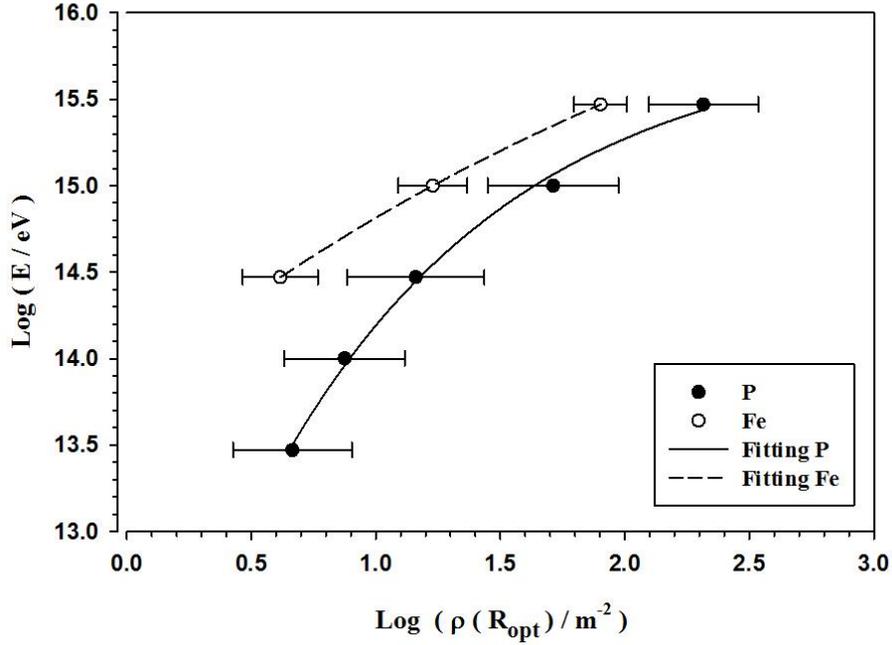

Fig.7: LogE in terms of Logρ ($R_{opt}$) for P and Fe initiated showers and resulting curve along them useing NKG distribution function.

There is no data point for Iron in low energies because density is very low and these showers cannot be recorded. The fitted energy estimation functions are:

$$LogE = 15.87 - 4.70\exp(-1.03 Log\rho(R_{opt})) \qquad (7)$$

for proton and

$$LogE = 17.32 - 3.50\exp(-0.34 Log\rho(R_{opt})) \qquad (8)$$

for Iron.

Fig. 7 also shows that light and heavy primary particles can be distinguished in the same energy.

## Verification of energy estimation function

To check validity of the energy estimation function, ρ($R_{opt}$) is calculated for a new set of vertical showers and substituted in equation 7 and 8 to estimate primary energy. Fig.8 demonstrates logarithm of the actual energy of showers versus of estimated energy by defined function. For easier comparison, bisectors are indicated in the figure (lines). The figure point out that the estimated energy by applied functions is more accurate in higher energies. (Number of showers used in different energy for each primary mass (P and Fe) is shown in the table next to the Fig.8)

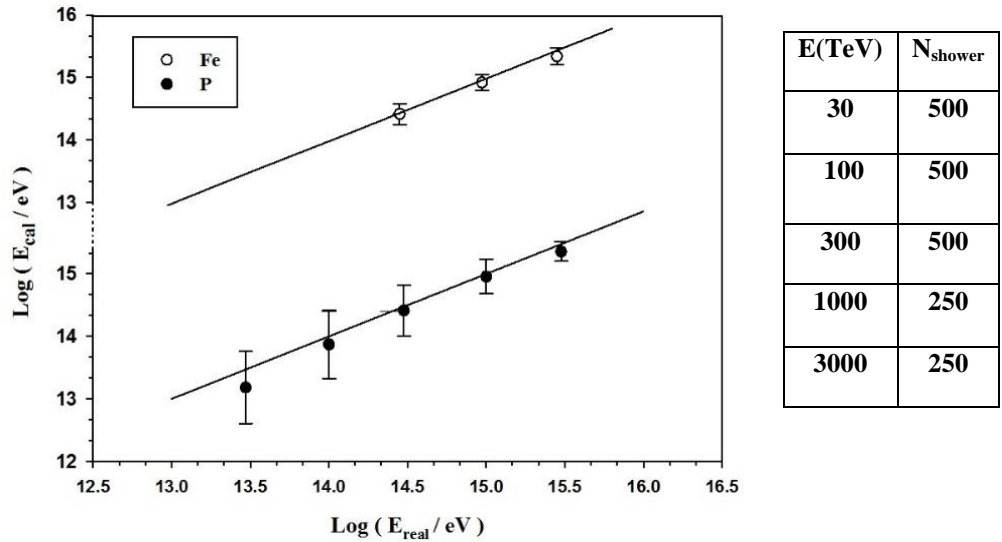

| E(TeV) | $N_{shower}$ |
|---|---|
| 30 | 500 |
| 100 | 500 |
| 300 | 500 |
| 1000 | 250 |
| 3000 | 250 |

Fig.8: Average logarithm of energy estimated by equations 7&8 versuse logarithm of actual energy used in CORSIKA input file for P and Fe. Note scale in vertical axes is restarted for Fe primary. Lines are Bisector for easier comparison as explained in the text.

## Conclusion

Optimum distance ($R_{opt}$) where the density calculated by lateral distribution function, has its minimum uncertainty explained via both analytical and numerical method. According to our results, these two methods lead to the same result for finding position of optimum distance. Therefore, results of this work have derived based on the analytical method only. In addition, using 3 different LDF models, it is shown that optimum distance is independent to LDF models. Then using analytical method and NKG function for LDF and considering trigger condition of 5 central detectors, $R_{opt}$ is found about 9±1m, for Alborz-I array.

This study shows that $R_{opt}$ has a little dependency to the characteristics of the primary particle like mass energy and zenith angle. Moreover, it is shown that $R_{opt}$ is thoroughly dependent to the layout and detector spacing. As shown in table 3 and 4 for 20 scintillation detector with the size of 50×50cm$^2$ the best $R_{opt}$ is achieved for Alborz-I array layout and its detector spacing (5 m).

Finally, estimation functions for energy was introduced for proton and Iron, as it is demonstrated in Fig.7 they can calculate energy in terms of density in $R_{opt}$ with good approximation especially in higher energies.

# Acknowledgement

The authors would like to thank Dr. Mahmoud Bahmanabadi for his valuable comments. We are also grateful to Ms. Saba Mortazavi moghaddam for her help and support.
# Reference

[1] S. P. Knurenko & A. Sabourov, The depth of maximum shower development and its fluctuations: cosmic ray mass composition at $E_0 \geq 10^{17}$ eV, *Astrophysics and Space Sciences Transactions* 7 (2011) 251-255.

[2] G. Rastegarzadeh & L. Rafezi, Energy, altitude, and mass dependence of steepness of the lateral distribution function of electrons and muons in extensive air showers, *Nuclear Instruments and Methods in Physics Research Section A* 763 (2014) 197-201.

[3] R. R. Prado et al., Interpretation of measurements of the number of muons in extensive air shower experiments, *Astroparticle Physics* 83 (2016) 40-52.

[4] V. de Souza et al., Shower size parameter as an estimator of extensive air shower energy in fluorescence telescopes, *Physical Review D* 73 (2006).

[5] G. Battistoni, et al., Monte Carlo study of the arrival time distribution of particles in extensive air showers in the energy range 1-100 TeV, *Astroparticle Physics* 9 (1998) 277-295.

[6] T. Antoni et al., Electron, muon, and hadron lateral distributions measured in air-showers by the KASCADE experiment, *Astroparticle Physics* 14 (2001) 245-260.

[7] A .M. Hillas et al., Measurement of primary energy of air showers in the presence of fluctuations, Proc. 12th ICRC, Hobart, Australia, 3(1971) 1001.

[8] D. Newton et al., The optimum distance at which to determine the size of a giant air shower, *Astroparticle Physics* 26 (2007) 414–419.

[9] S .Abdollahi et al., Alborz-I array: A simulation on performance and properties of the array around the knee of the cosmic ray spectrum, *Astroparticle Physics* 76 (2016) 1–8.

[10] D. M. Edge et al., The cosmic ray spectrum at energies above $10^{17}$ eV, *Physics A* 6 (1973) 1612.

[11] A.Tapia et al, The lateral shower age parameter as an estimator of chemical composition, Proc. 33th ICRC, Rio de Janeiro, Brazil, (2013).

[12] P.K.F.Grieder, Extensive Air Showers, Springer, vol1, (2010).